\documentclass[preprint,showpacs,
preprintnumbers,amsmath,amssymb]{revtex4}
\usepackage{graphicx}
\usepackage{dcolumn}
\usepackage{bm}
\usepackage[usenames]{color}

\def\F{\mbox{F}}
\def\E{\mbox{E}}
\def\L{{\cal L}}
\def\H{\mbox{H}}

\def\exp{\mbox{exp}}
\def\tg{\widetilde{\gamma}}

\begin{document}
\title{An alternative factorization of the quantum harmonic oscillator and two-parameter family of self-adjoint operators}
\author{R Arcos-Olalla}
\email{olalla@fisica.ugto.mx}
\affiliation{Departamento de F\'{\i}sica, DCI Campus Le\'on, Universidad de Guanajuato\\ Apdo. Postal E143, 37150 Le\'on, Gto., Mexico}
\author{Marco A. Reyes}
\email{marco@fisica.ugto.mx}
\affiliation{Departamento de F\'{\i}sica, DCI Campus Le\'on, Universidad de Guanajuato\\ Apdo. Postal E143, 37150 Le\'on, Gto., Mexico}
\author{H.C. Rosu}
\email{hcr@ipicyt.edu.mx}
\affiliation{IPICYT, Instituto Potosino de Investigacion Cientifica y Tecnologica,\\Apdo Postal 3-74 Tangamanga, 78231 San Luis Potos\'{\i}, S.L.P., Mexico}
\date{September 2012}
\begin{abstract}
We introduce an alternative factorization of the Hamiltonian of the quantum harmonic oscillator which leads to a two-parameter self-adjoint operator from which the standard harmonic oscillator, the one-parameter oscillators introduced by Mielnik, and the Hermite operator are obtained in certain limits of the parameters. In addition, a single Bernoulli-type parameter factorization which is different of the one introduced by M. A. Reyes, H. C. Rosu, and M. R. Guti\'errez, Phys. Lett. A 375 (2011) 2145 is briefly discussed in the final part of this work.
\end{abstract}

\begin{center}
Physics Letters A\\ 2012
\end{center}

\maketitle


\section{Review of our previous work}\label{s1}

In \cite{1}, the factorization of the Hamiltonian of the quantum harmonic oscillator has been effected
by means of the pair of non-mutually adjoint operators
\begin{subequations}
\begin{align}\label{bmenos}
 B^- &= \frac{1}{\sqrt{2}}\left(\alpha^{-1}(x)\frac{d}{dx}+\beta(x)\right),\\ \label{bmas}
B^+ &= \frac{1}{\sqrt{2}}\left(-\alpha(x)\frac{d}{dx}+\beta(x)\right).
\end{align}\label{Bs}
\end{subequations}
We notice that these operators coincide with the standard creation and annihilation operators $a$ and $a^*$
when $\alpha(x)=1$ and $\beta(x)=x$ and with the operators introduced by Mielnik \cite{M} $b$ and $b^*$
when $\alpha(x)=1$ and $\beta(x)=x+\frac{\F'(x)}{\gamma_1+\F(x)}$ where $\F(x)=\int_0^x e^{-t^2}dt$ is related to the error function $\E(x)$ through
$\F(x)=\frac{\sqrt{\pi}}{2}\E(x)$
while $\gamma_1$ is the integration constant parameter occurring through the integration of the Riccati equation.
The factorization $B^-B^+=H+\frac{1}{2}$ corresponding to the eigenvalue problem $H\psi=\lambda\psi$ of the quantum harmonic oscillator is effectively written as
\begin{equation}\label{op_prod}
 B^-B^+ \equiv \frac{1}{2}\left[ -\frac{d^2}{dx^2}-\left( \frac{\alpha'-\beta}{\alpha}+\alpha\beta \right)
\frac{d}{dx}+\frac{\beta'}{\alpha}+\beta^2\right]= \frac{1}{2}\left( -\frac{d^2}{dx^2}+x^2 +1\right),
\end{equation}
so that $\alpha(x)$ and $\beta(x)$ should fulfill the following coupled equations
\begin{subequations}
\begin{align}\label{a}
 & \alpha'+\beta\alpha^2-\beta = 0,\\ \label{b}
 & \beta'+\alpha\beta^2 = \left(1+x^2\right)\alpha.
\end{align}\end{subequations}
By decoupling these equations one gets the Riccati equation:
\begin{equation}\label{ed0}
 \frac{d}{dx}\left(\frac{\beta}{\alpha}\right)+\left(\frac{\beta}{\alpha}\right)^2 = 1+x^2.
\end{equation}
In \cite{1}, a particular solution has been obtained by taking $\beta(x)/\alpha(x)=x$ which leads to the following simple form of the functions
$\alpha$ and $\beta$
\begin{equation}
 \alpha_{\delta}(x)=\frac{1}{\sqrt{1+\delta e^{-x^2}}}, \hspace{1.5cm} \beta_{\delta}(x)=\frac{x}{\sqrt{1+\delta e^{-x^2}}},
\hspace{1.5cm}\delta=\mbox{constant}.     \label{ranf}
\end{equation}
To avoid singularities, the parameter $\delta$ should be in the range $-1<\delta<\infty$.

The goal of this Letter is to obtain a complete general solution of the above Riccati equation (\ref{ed0}) which occurs in the alternative factorization based on operators which are not mutually adjoint. This solution enables one, as shown in section II, to construct a two-parameter family of self adjoint operators from which the standard harmonic oscillator, the one-parameter oscillators introduced by Mielnik, and the Hermite operator can be obtained in particular cases.  In section III, we discuss another possibility of factorization of the quantum harmonic oscillator, apart from the one addressed in the previous section. Finally, we summarize our work in section IV.


\section{Two-parameter self-adjoint operator from alternative factorization}\label{fa}

Here, we are interested in the most general solution of (\ref{ed0}). Since this is just the Riccati equation for the standard harmonic oscillator, the solution is well known
\begin{equation}\label{beta0}
\frac{\beta}{\alpha} = x + \frac{e^{-x^2}}{\gamma_1+\F(x)}~.
\end{equation}
Using $\beta(x)$ from (\ref{beta0}) in (\ref{a}) we get 
\begin{equation}
 \alpha'+\alpha\left(\alpha^2-1\right)\left(x + \frac{e^{-x^2}}{\gamma_1+\F(x)}\right) = 0~,
\end{equation}
which is a Bernoulli equation that can be solved by the method of separation of variables. The general solution reads:
\begin{equation}
 \alpha_{\gamma_1\gamma_2}(x)=\pm\frac{1}{\sqrt{1-\gamma_2\left(\:\gamma_1+\F(x)\:\right)^{-2}e^{-x^2}}}~,
\end{equation}
where $\gamma_2$ is a Bernoulli integration constant.
In the following, we choose the solution with the plus sign because due to eq.~(\ref{beta0}) the other sign cancels out in operators product (\ref{op_prod}). Thus:
\begin{equation}\label{alpha0}
%
\alpha_{\gamma_1\gamma_2}(x)=\frac{|\gamma_1+\F(x)|}{\sqrt{\left(\, \gamma_1+\F(x) \,\right)^2-\gamma_2\,e^{-x^2}}}~,
\end{equation}
and
\begin{equation}
 \beta_{\gamma_1\gamma_2}(x)=\frac{x + \left(\gamma_1+\F(x)\right)^{-1} \, e^{-x^2}}
 {\sqrt{1-\gamma_2\left(\:\gamma_1+\F(x)\:\right)^{-2}e^{-x^2}}}~.
\label{beta1}
\end{equation}
The subindices could be occasionally omitted henceforth.
Since the operators $B^-$ and $B^+$ define the Hamiltonian, they should be well behaved for any $x$. This depends on the value of the parameters $\gamma_1$ and $\gamma_2$ in $\alpha_{\gamma_1\gamma_2}(x)$ and $\beta_{\gamma_1\gamma_2}(x)$ as discussed further.
It is also worth noting that the product $\left(\frac{1}{\alpha}\frac{d}{dx} \right)\left( \alpha\frac{d}{dx} \right)$ in $B^-B^+$ looks similar
to the Hartle-Hawking factor ordering \cite{hartle} in SUSY quantum cosmology \cite{socorro}.

From the expressions of the operators (\ref{Bs}) and of (\ref{alpha0}) and (\ref{beta1}), it is clear that we have to determine the values of the parameters in such a way that $\alpha_{\gamma_1\gamma_2}(x)$  is not either zero or not defined.
From the numerator of (\ref{alpha0}), and taking into account that
$-\sqrt{\pi}/2<\F(x)<\sqrt{\pi}/2$, the condition $|\gamma_1|>\frac{\sqrt{\pi}}{2}$ 
does not allow $\alpha_{\gamma_1\gamma_2}(x)$ to become zero. This is exactly Mielnik's condition on his parameter $\gamma$.
In addition, if $\gamma_2\leq 0$ the denominator cannot be zero.
On the other hand, to see what happens in the case $\gamma_2>0$ is not as simple because the resulting inequality is transcendental.
The fact that $(\gamma_1+\F(x))^2-\gamma_2e^{-x^2}>0$ leads to:
$(\gamma_1+\F(x))^2>\gamma_2e^{-x^2}$.
Since $\gamma_2e^{-x^2}\leq\gamma_2$ for all $x$ we will consider only those values of $\gamma_1$ for which $(\gamma_1+\F(x))^2>\gamma_2$.
Taking the maximum and minimum allowed values of the error function in this inequality we obtain 
$ \left(\gamma_1+\frac{\sqrt{\pi}}{2}\right)^2>\gamma_2$ and $ \left(\gamma_1-\frac{\sqrt{\pi}}{2}\right)^2>\gamma_2$, respectively.
Their product reads
$ \left|\gamma_1^2-\frac{\pi}{4}\right|>\gamma_2$, 
where we have eliminated the modulus in the right hand side because we consider only the case $\gamma_2>0$.
However, a numerical approach shows that an inequality which better forbids the appearance of singularities in the transcendental equation is
$\gamma_2<\gamma_1^2-1$, and therefore, the two parameters must satisfy the inequalities $|\gamma_1|>\sqrt{\pi}/2$, and $\gamma_2<\gamma_1^2-1$ or $\gamma_2\leq0$.

The various factorizations can be now obtained from the general factorization as follows:
The functions $\alpha_{\delta}(x)$ and $\beta_{\delta}(x)$ found in \cite{1} are obtained if we impose the conditions:
$\frac{\gamma_2}{\gamma_1^2}=-\delta$ and $|\gamma_1|\rightarrow\infty$.       
For the case of the factorization introduced by Mielnik it is enough to take $\gamma_2=0$ and $\gamma_1=\gamma$.
Finally, the case of the standard factorization is obtained if we have $\alpha_{\gamma_1\gamma_2}(x)=1$ and $\beta_{\gamma_1\gamma_2}(x)=x$, such that $\gamma_2=0$ and $|\gamma_1|$ tends to infinity.

\begin{figure}[H]\begin{center}\includegraphics[width=0.8\textwidth]{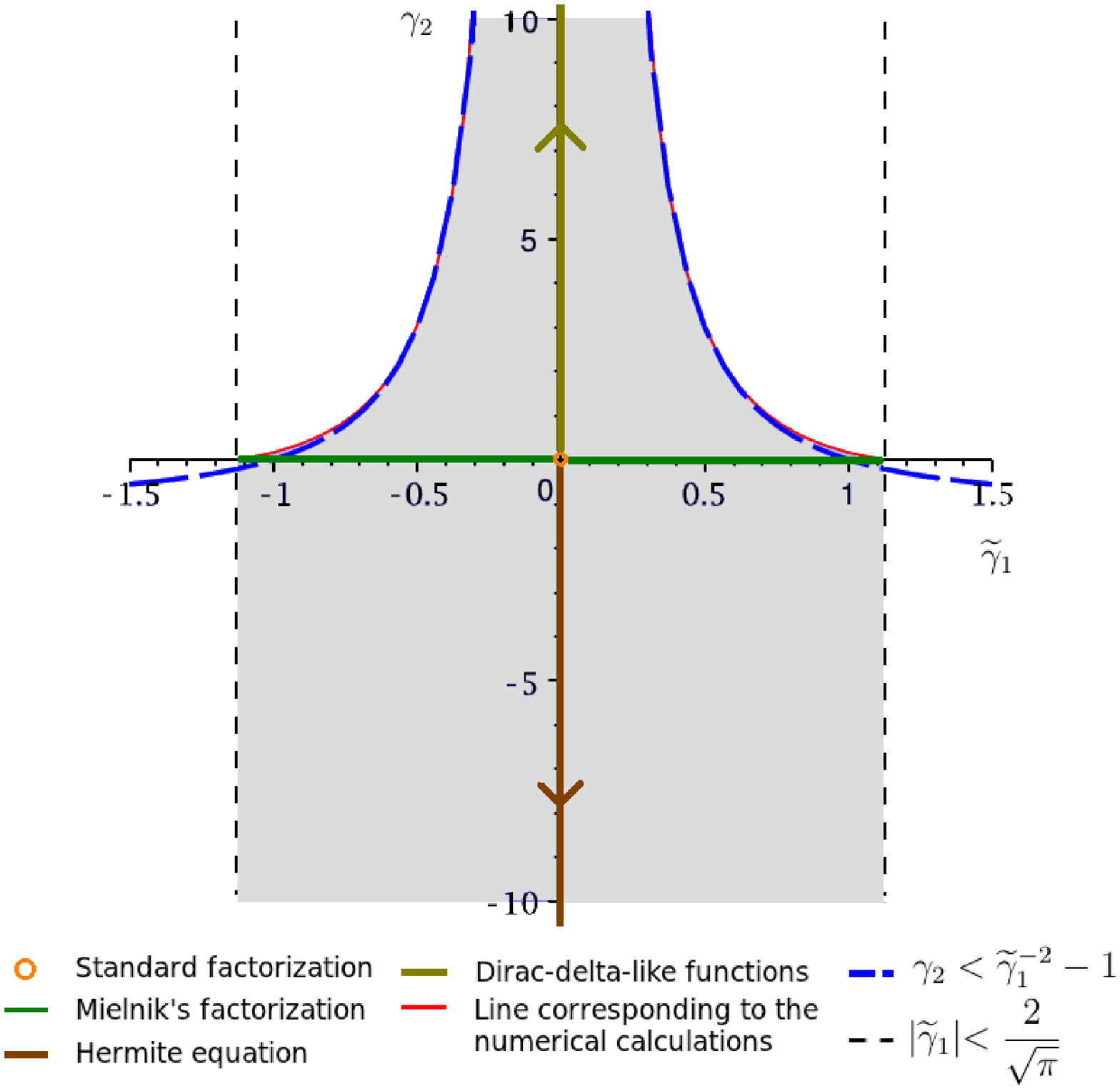}
\caption{\label{par4} Plot of the allowed region in the parameter space;
in particular, we emphasize the values of the parameters that lead to the factorizations studied in this work: the vertical axis corresponds to the factorization obtained in \cite{1}, whereas the horizontal axis corresponds to Mielnik's factorization.
The Hermite equation and the delta-like singular $\H_0^{\gamma_{1,2}}$ are found in the extremes of the vertical axis.
}\end{center}\end{figure}

In Figure \ref{par4}, we can see the geometric positions in the parameter space defining the various factorizations, with the $x$-axis defined by
$\tilde{\gamma_1}=1/\gamma_1$ and the $y$-axis defined by $\gamma_2$. The reason to use the inverse of the parameter introduced by Mielnik is to provide a more convenient display of the factorizations in the two-parameter space, but perhaps it is interesting to note that it has been used before in other SUSY developments \cite{rr98,berezovoj}. Thus, in fig.~\ref{par4} one gets:
(i) The standard factorization $\gamma_{1}^{-1}\rightarrow\pm\infty$ or $\tg_1\rightarrow0$, $\gamma_2=0$ corresponds to the origin in this new parameter space; 
(ii) The factorization introduced by Mielnik: $\gamma_2\equiv 0$, on the horizontal axis between $(-\frac{2}{\sqrt{\pi}},
\frac{2}{\sqrt{\pi}})$;  
(iii) The factorizations with the help of operators which are not mutually adjoint in \cite{1} are encountered when one moves along the vertical axis.
The case $\delta\rightarrow-1$ is obtained when $\gamma_2\rightarrow+\infty$. If $\delta>0$, we move along the negative vertical axis with the Hermite equations corresponding to the limit $\gamma_2\rightarrow-\infty$.

\begin{figure}[H]\begin{center}\includegraphics[width=0.7\textwidth]{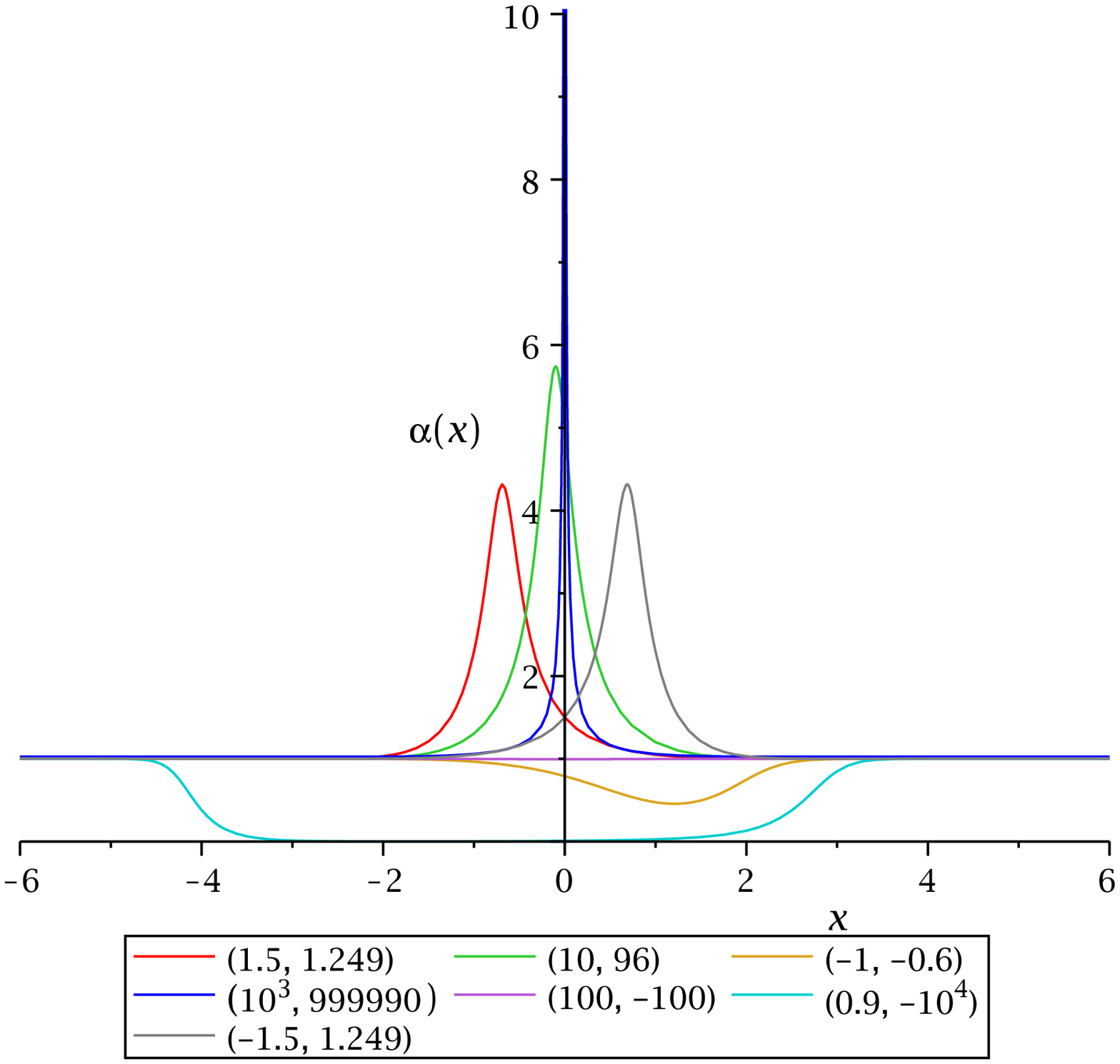}
\caption{\label{falpha} Plots of the functions $\alpha(x)$ for different values of the parameters ($\gamma_1,\gamma_2$).
The highest peak in this panel, corresponding to the parameter pair $(\gamma_1,\gamma_1-10)$ for $\gamma_1=10^3$, is also of the smallest width.
}
\end{center}\end{figure}

\begin{figure}[H]\begin{center}\includegraphics[width=0.7\textwidth]{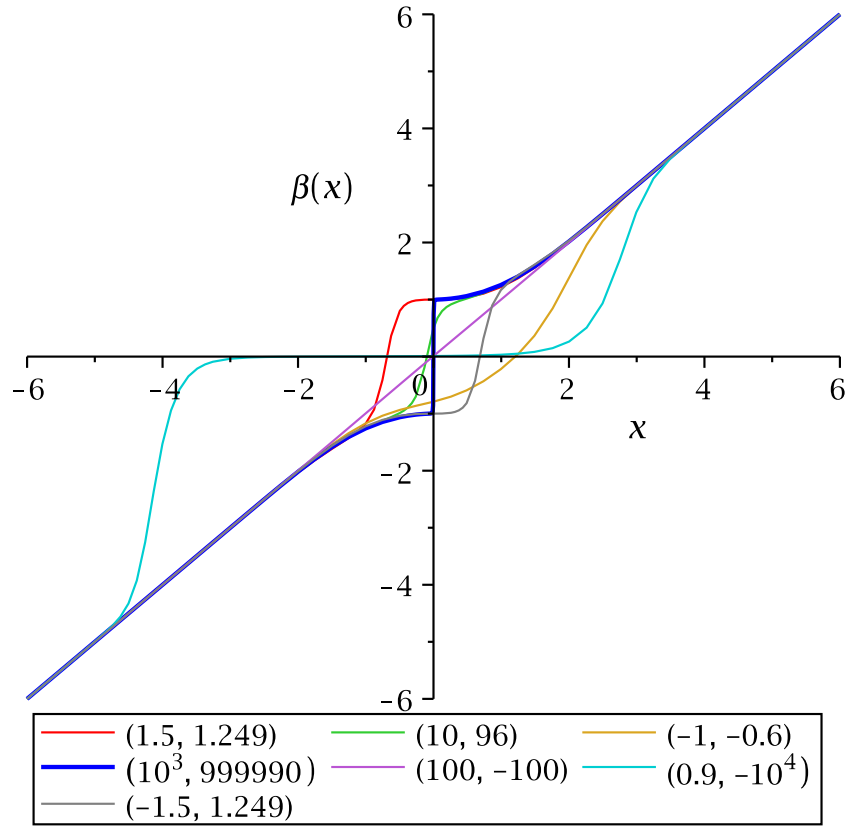}
\caption{\label{gbeta} Plots of $\beta(x)$ for the same values of the parameters
as used in Fig. \ref{falpha}. The difference from the standard $\beta(x)=x$ plot shrinks at high values of the parameters.
}
\end{center}\end{figure}

We also display several graphics of $\alpha_{\gamma_1 \gamma_2}(x)$ with the goal to show how it changes as a function of the parameters
$\gamma_1$ and $\gamma_2$ in Fig.~\ref{falpha}. 
In the case of the curves that lie above the line $\alpha_{\gamma_1\gamma_2}=1$, one can see that the peaks pass from the first quadrant to the second one or viceversa when the sign of $\gamma_1$ is changed, see, e.g., the curves corresponding to the pairs of parameters (1.5, 1.249) and (-1.5, 1.249). If we take values very close to the dashed (blue) curve in
Fig.~\ref{par4}, then the peak of $\alpha_{\gamma_1\gamma_2}(x)$ becomes bigger at higher values of the parameters. All these curves have been scaled in order to have all of them in a single plot. Notice that for values close to $\gamma_1=\pm\sqrt{\pi}/2$, for $\gamma_2<0$,
$\alpha_{\gamma_1\gamma_2}(x)$ is below the line $\alpha_{\gamma_1\gamma_2}(x)=1$.
On the other hand, some plots of $\beta_{\gamma_1\gamma_2}(x)$ for the same values of the parameters are shown in Fig. \ref{gbeta}. Since asymptotically $\alpha_{\gamma_1 \gamma_2}(x)\rightarrow 1$ and the second term in the right hand side of eq.~(\ref{beta0}) goes to zero, we can see the asymptotic bisectrix behaviour of this coefficient.

\begin{figure}[H]\begin{center}\includegraphics[width=0.65\textwidth]{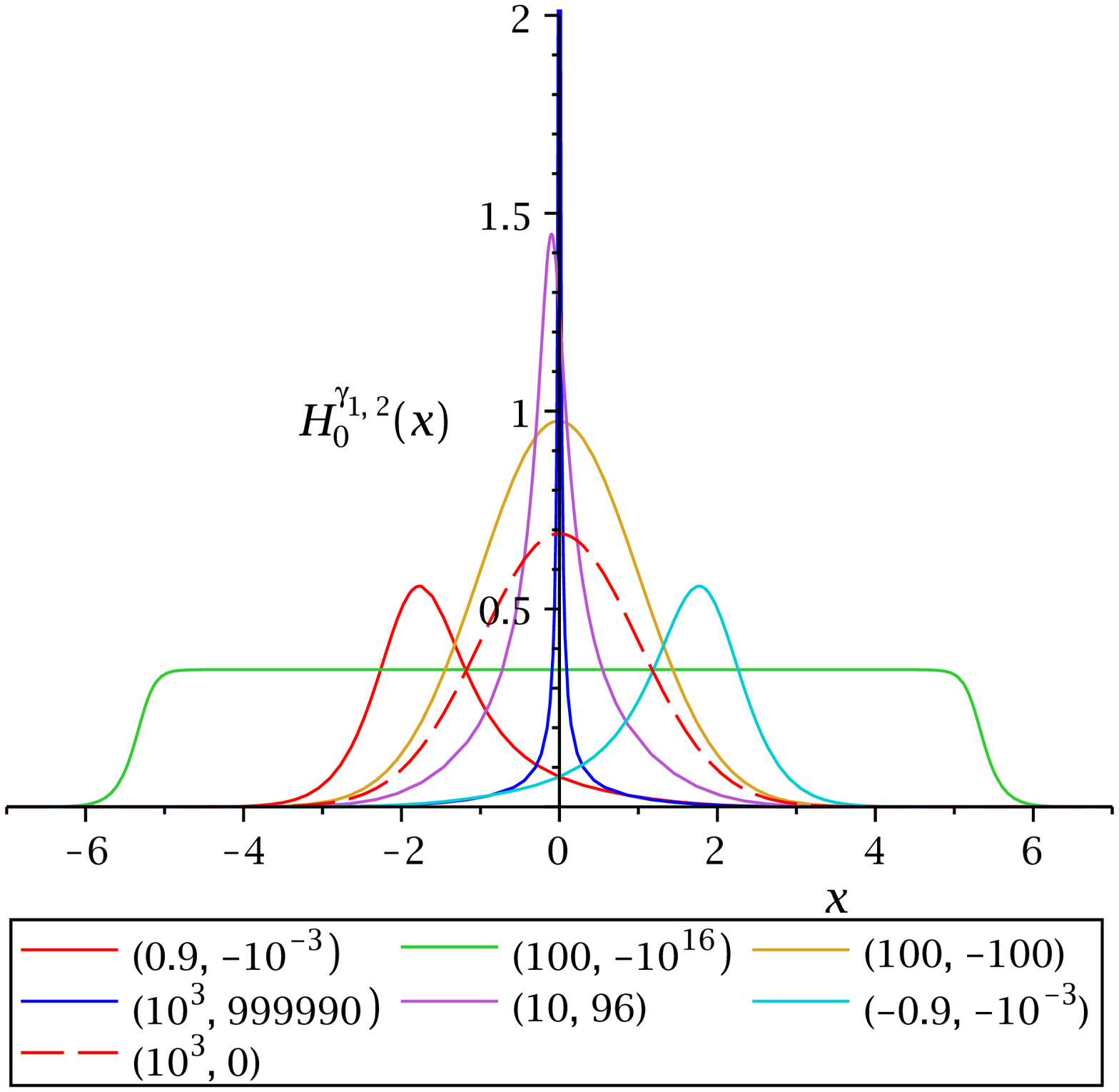}
\caption{\label{h0}The function $\H_0^{\gamma_{1,2}}(x)$ for different values of the pair of parameters. Notice the symmetry of the
red and cyan curves, which differ only in the sign of $\gamma_1$. For all non-zero $\gamma_2$, these functions are not Gaussian.}
\end{center}\end{figure}

We now deal with the reversed product operator $B^+B^-$. 
It is more convenient to add $\frac{1}{2}$ to relate it to the Dirac factorization of the harmonic oscillator.
Then, $\widetilde\L_{\gamma_{1,2}}=B^+B^-+\frac{1}{2}$ leads to:
%
%
\begin{eqnarray}
 \widetilde\L_{\gamma_{1,2}} &\equiv&\frac{1}{2}\left[-\frac{d^2}{dx^2}-\frac{2\gamma_2\alpha\beta e^{-x^2}}{\left(\gamma_1+\F(x)\right)^{2}}\:\,\frac{d}{dx}
+\left( 1 + \alpha^2 \right)\beta^2-\left( 1 + x^2 \right)\alpha^2+1\right]  . \label{operador0}
\end{eqnarray}
Introducing the functions: 
\begin{equation}
 \H^{\gamma_{1,2}}_{n+1}(x) = B^+\psi_n(x), \label{def_func}
\end{equation}
where $\psi_n(x)$ are the eigenfunctions of the harmonic oscillator and applying the $\widetilde\L_{\gamma_{1,2}}$ operator
\begin{equation}
 \widetilde\L_{\gamma_{1,2}} \H^{\gamma_{1,2}}_{n+1} \equiv \left(B^+B^-+\frac{1}{2}\right)\left(B^+\psi_n\right)\equiv
B^+\left(B^-B^+ +\frac{1}{2}\right)\psi_n
= \left( \lambda_{n}+1 \right)\H^{\gamma_{1,2}}_{n+1}~,   \label{eigL0}
\end{equation}
where $\lambda_n=n+\frac{1}{2}$ are the harmonic oscillator eigenvalues, one can see that they are the eigenfunctions of the $\widetilde\L_{\gamma_{1,2}}$ operator, but without including
the $\H_0^{\gamma_{1,2}}$ function. However, the latter one can be introduced in the usual SUSY manner
asking that $\widetilde\L_{\gamma_{1,2}}\H_0^{\gamma_{1,2}}=\lambda_0\H_0^{\gamma_{1,2}}$, or
\begin{equation}
\left( B^+B^-+\dfrac{1}{2} \right)\H_0^{\gamma_{1,2}}=\dfrac{1}{2}\H_0^{\gamma_{1,2}}~.
\end{equation}
This leads to:
\begin{equation}
 B^+B^-\H_0^{\gamma_{1,2}}=0,
\end{equation}
which requires
\begin{equation}
 B^-\H_0^{\gamma_{1,2}} = \frac{1}{\sqrt{2}}\left[\frac{1}{\alpha_{\gamma_1\gamma_2}(x)}\:\frac{d}{dx}+
\alpha_{\gamma_1\gamma_2}(x)\left(x + \frac{e^{-x^2}}{\gamma_1+\F(x)}\right)\right]\H_0^{\gamma_{1,2}} = 0~.
\end{equation}
The solution of this equation is:
\begin{equation}
 \H_0^{\gamma_{1,2}}(x) = \frac{\psi_0(x)}{\sqrt{\left( \gamma_1+\F(x) \right)^2-\gamma_2e^{-x^2}}}=\alpha_{\gamma_1\gamma_2}(x)\psi_{0M}(x)~,
\end{equation}
where $\psi_{0M}(x)=\dfrac{\psi_0}{|\gamma_1+\F(x)|}$ is the modulated zero mode obtained by Mielnik.
In fig.~\ref{h0} various plots of $\H_0^{\gamma_{1,2}}(x)$ are displayed for some values of the parameters.
When $\gamma_2\rightarrow+\infty$, these functions become more and more singular.

Since in general any second-order differential equation $Pu''+Qu'+Ru+\lambda u=0$ can be transformed to the self-adjoint form $\frac{d}{dx}\left(pu'\right)+qu+\lambda\omega(x)u=0$, multiplying it by the factor $(1/P)\mbox{exp}\left(\int^x(Q/P)dx\right)$,
we find in our case that this factor is $-2\alpha^{-2}(x)$ which leads to the following eigenvalue equation:
\begin{equation}
 \L_{\gamma_{1,2}}\H_n^{\gamma_{1,2}}(x)+\lambda_n\omega_{\gamma_{1,2}}(x)\H_n^{\gamma_{1,2}}(x)=0, \label{ec_eigenv}
\end{equation}
where
%
\begin{eqnarray}
 \L_{\gamma_{1,2}}&=&
\frac{d}{dx}\left[\left( 1-\frac{\gamma_2e^{-x^2}}{\left( \,\gamma_1+\F(x)\, \right)^2} \right)\frac{d}{dx}\right]
+ x^2-\bigg[ \left( 1-\frac{\gamma_2e^{-x^2}}{\left( \,\gamma_1+\F(x)\, \right)^2} \right)^{-1}+1 \bigg]
\beta_M^2
\nonumber\\
&&
+ \frac{\gamma_2e^{-x^2}}{\left( \,\gamma_1+\F(x)\, \right)^2} ~,
\qquad \qquad \qquad  \beta_M = x+\frac{\F'(x)}{\gamma_1+\F(x)}~,
\label{o_autoad}
\end{eqnarray}
is a new self-adjoint harmonic oscillator operator with $\omega_{\gamma_{1,2}}(x)=2\alpha^{-2}_{\gamma_1\gamma_2}(x)$
as weight function, which according to the general Sturm-Liouville theory should be strictly positive except possibly at isolate points where
$ \omega_{\gamma_{1,2}}(x)=0$ \cite{arfken}.

A convenient expression for the eigenfunctions of the operator $\L_{\gamma_{1,2}}$ can be obtained by writing $B^+$ in terms of
$a$ and $a^*$ because $B^+$ applied to the functions $\psi_n$ convert them into the eigenfunctions $\H_n^{\gamma_{1,2}}(x)$ of $\L_{\gamma_{1,2}}$: 
\begin{equation}
 B^+=\alpha_{\gamma_1\gamma_2}(x)\left(a^*+ \frac{e^{-x^2}}{\gamma_1+\F(x)}\right)~.
\end{equation}
This implies the following relationship between the two sets of eigenfunctions:
\begin{equation}\label{autof}
 \H_{n+1}^{\gamma_{1,2}}(x)= \alpha_{\gamma_1\gamma_2}(x)\left[\sqrt{n+1}\:\psi_{n+1}(x)+\frac{e^{-x^2}}{\gamma_1+\F(x)}\:\psi_n(x)\right].
\end{equation}
These eigenfunctions are orthogonal because their construction is performed according to the Sturm-Liouville theory and correspond to the same equidistant harmonic oscillator spectrum because $ \L_{\gamma_{1,2}}$ is isospectral to the harmonic oscillator Hamiltonian (\ref{ec_eigenv}).

\begin{figure}[H]\begin{center}\includegraphics[width=0.63\textwidth]{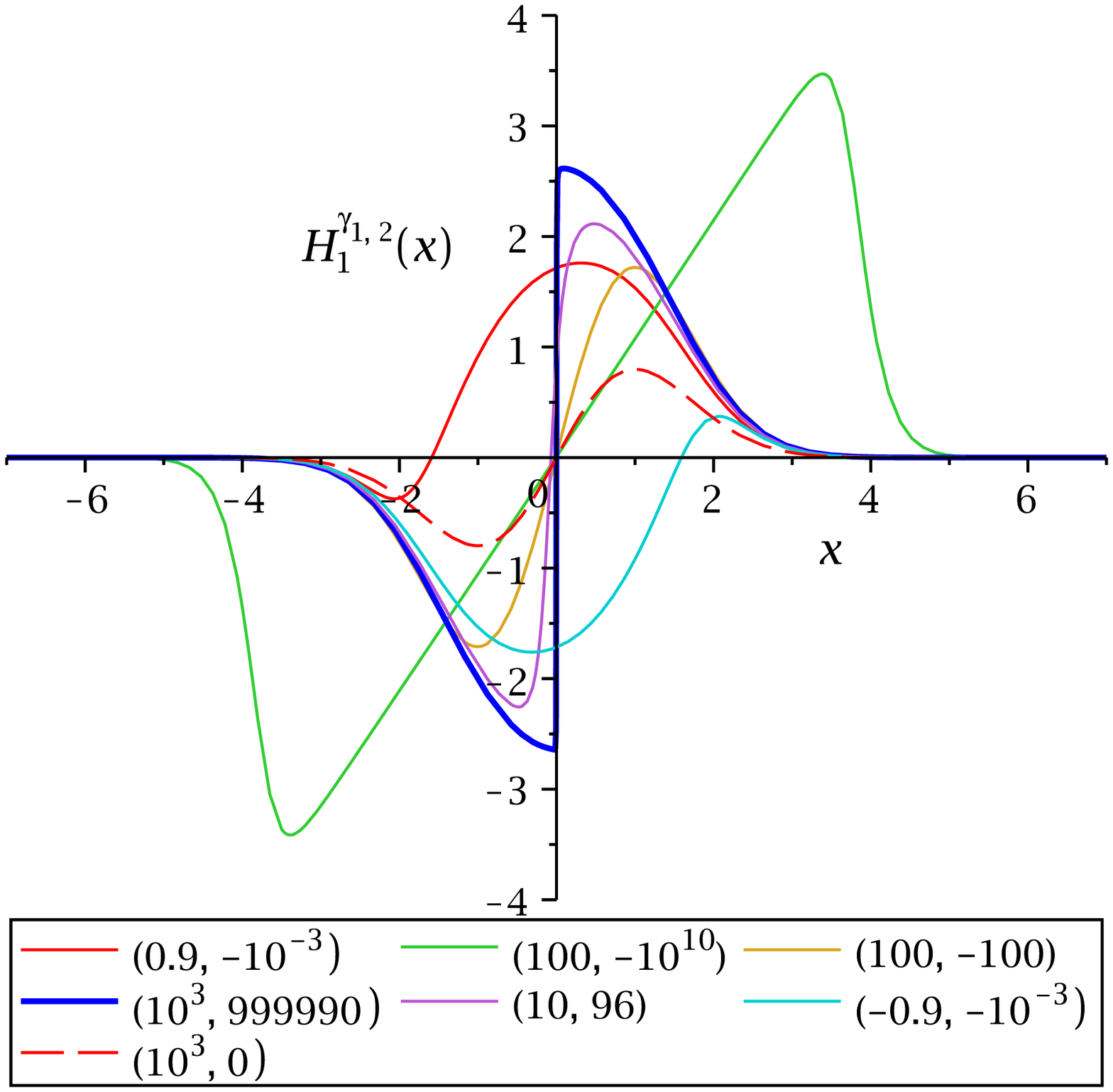}
\caption{\label{h1} The functions $\H_1^{\gamma_{1,2}}(x)$ for different values of the parameter pairs.}
\end{center}\end{figure}

Figures \ref{h0} and \ref{h1} display the plots of the ground-state eigenfunction $\H_{0}^{\gamma_{1,2}}(x)$ and first excited state eigenfunction $\H_{1}^{\gamma_{1,2}}(x)$ for several representative cases of the parameters $\gamma_1$ and $\gamma_2$.
From the figure \ref{h0} one can infer that $\gamma_1$ acts as a shift parameter for the maximum of the wavefunction along the $x$-axis.
The physical interpretation of $\gamma_2$ is not as easy and clear but it has to do with the distortion of the shape of the eigenfunctions.
\begin{figure}[H]\begin{center}\includegraphics[width=0.62\textwidth]{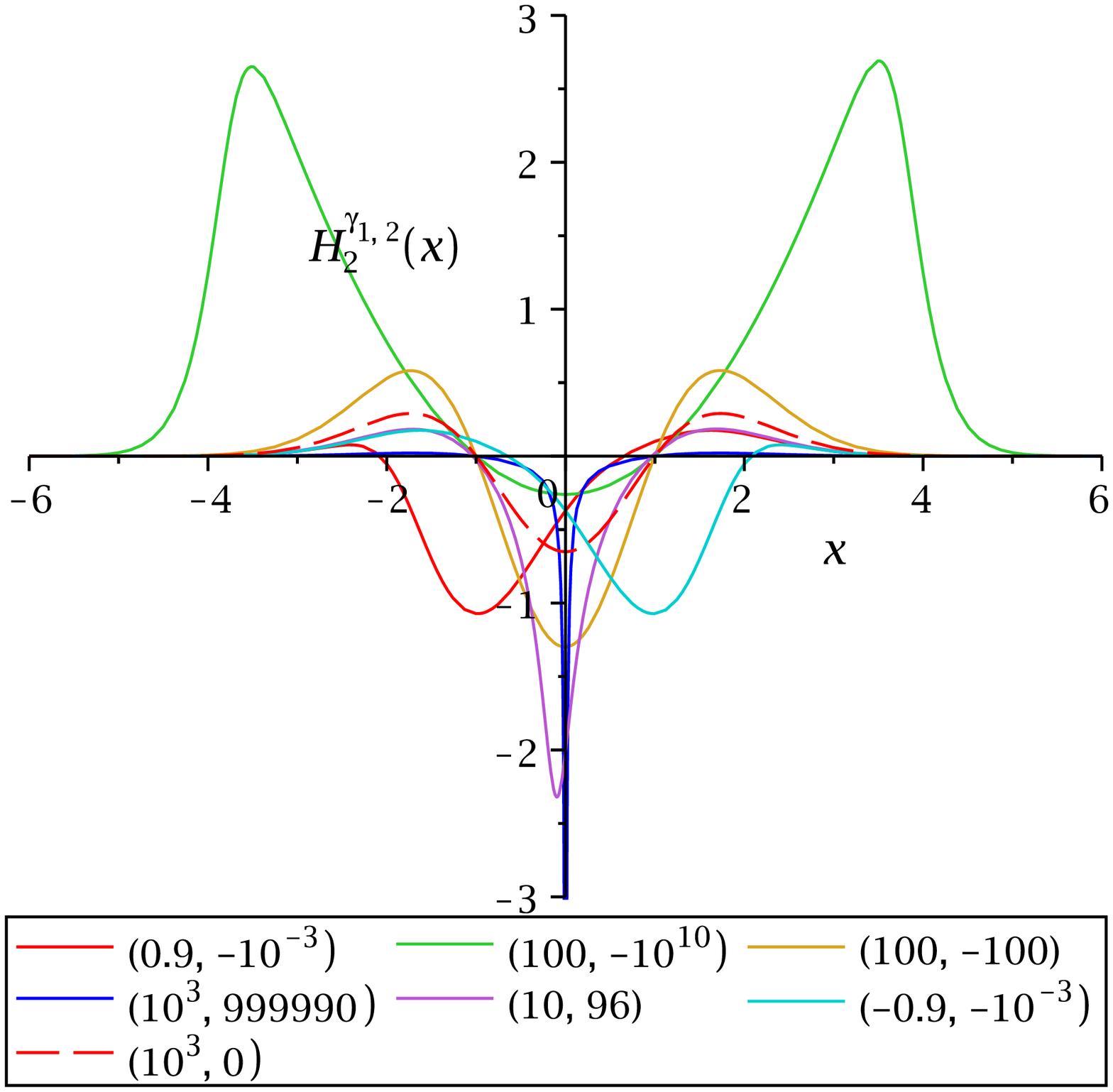}
\caption{\label{h2} Same as in the previous figure but for the functions $\H_2^{\gamma_{1,2}}(x)$. Figure that will not be included in the published version.}
\end{center}\end{figure}
\begin{figure}[H]\begin{center}\includegraphics[width=0.7\textwidth]{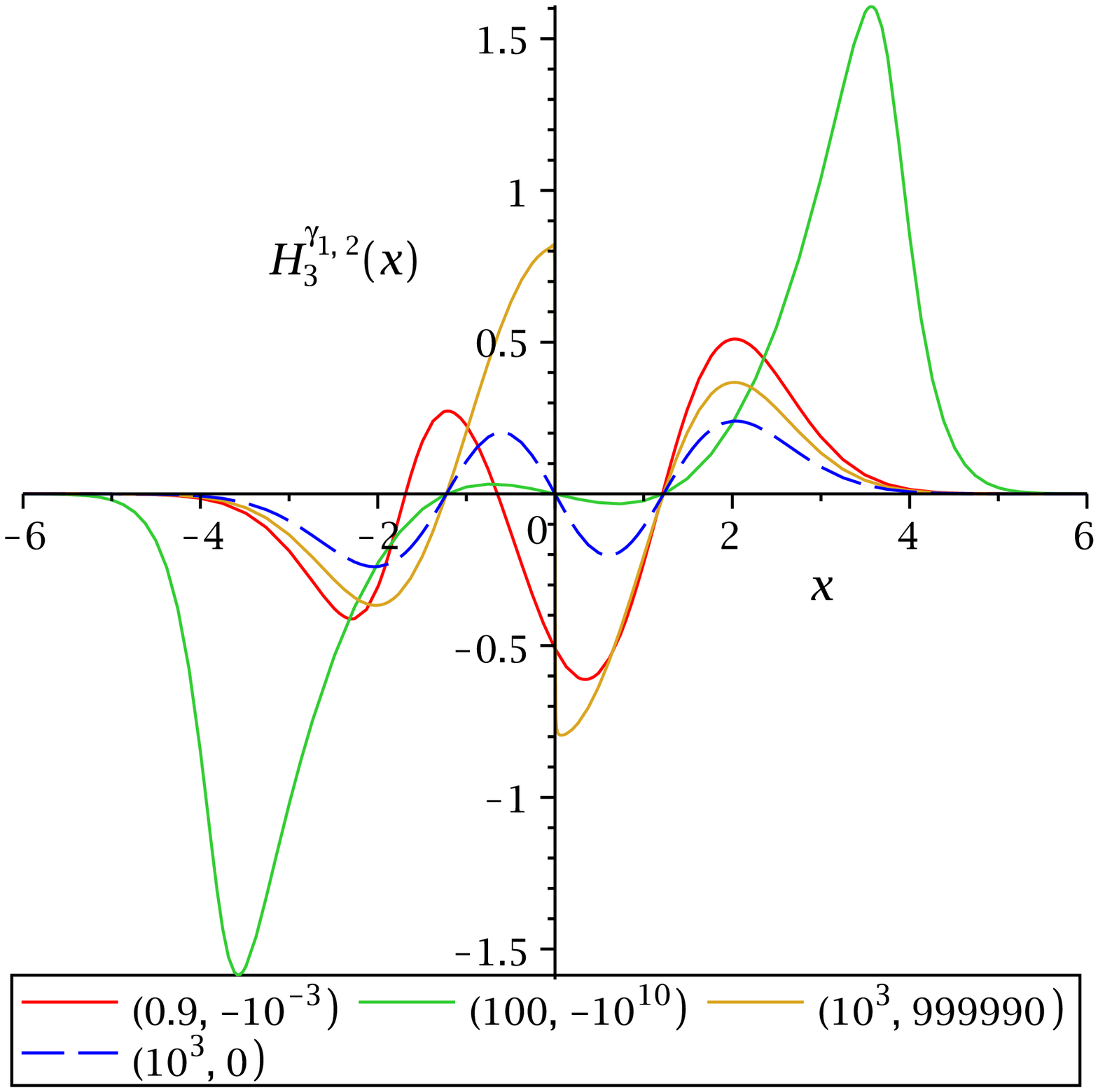}
\caption{\label{h3} Same as in the previous figures but for the functions $\H_3^{\gamma_{1,2}}(x)$. Figure that will not be included in the published version.}
\end{center}\end{figure}

It is now possible to comment on why in Fig.~\ref{par4}, we indicate a region for the Hermite equation.
The reason is the following. Assume that $|\gamma_1|\rightarrow\infty$ and at the same time the quotient
$-\gamma_2/\gamma_1^2\equiv\delta$ is big but still a finite number, then the self-adjoint operator (\ref{o_autoad}) takes the form:
\begin{equation}
 \L_{\gamma_{1,2}}\rightarrow\delta e^{-x^2}\left( \frac{d^2}{dx^2}-2x\frac{d}{dx}-1 \right),
\end{equation}
and moreover
\begin{equation}
 \lambda_n\omega_{\gamma_{1,2}}(x)\rightarrow\left( n+\frac{1}{2} \right)2\delta e^{-x^2}=\left( 2n+1 \right)\delta e^{-x^2}.
\end{equation}
On the other hand,
\begin{equation}
 \H_n^{\gamma_{1,2}}(x)\rightarrow \sqrt{n}\delta^{-\frac{1}{2}}\H_n(x)~,
\end{equation}
where $\H_n(x)$ are the Hermite polynomials. Therefore, in this approximation, eq.~(\ref{ec_eigenv}) turns into
\begin{equation}
 \sqrt{n}\delta^{\frac{1}{2}}\left( \frac{d^2\H_n}{dx^2}-2x\frac{d\H_n}{dx}+2n\H_n \right)=0,
\end{equation}
which is Hermite's equation. The way in which the functions $\H_n^{\gamma_{1,2}}(x)$ for $n=0$ and $n=1$ tend to the Hermite polynomials can be seen
in figures.~\ref{h0} and \ref{h1}.

Similarly to the Mielnik case, the functions $\H_{n}^{\gamma_{1,2}}(x)$ do not admit first-order creation operators. To show this fact suppose that the creation and annihilation operators for $\H_{n}^{\gamma_{1,2}}(x)$
are $A^+$ and $A$, respectively. Then one has:
\begin{equation}
 A^+\H_{n}^{\gamma_{1,2}}(x)=c_1\H_{n+1}^{\gamma_{1,2}}(x), \qquad A\:\H_{n}^{\gamma_{1,2}}(x)=c_2\H_{n-1}^{\gamma_{1,2}}(x),
\end{equation}
where $c_1$ and $c_2$ are constants. Using (\ref{def_func}) one can realize that applying the operator $B^-$ we get:
\begin{equation}
 B^-\H_{n}^{\gamma_{1,2}}(x)=B^-B^+\psi_{n-1}(x)=\left( H+\frac{1}{2} \right)\psi_{n-1}(x)=n\:\psi_{n-1}(x).
\label{ec_usar}
\end{equation}
Applying now the standard creation operator $a^{*}$, leads to:
\begin{equation}
 a^*B^-\H_{n}^{\gamma_{1,2}}(x)=na^*\psi_{n-1}(x)=n\sqrt{n}\:\psi_n(x).
\end{equation}
Finally, by applying $B^+$ we go from $\psi_n(x)$ to $\H_{n+1}^{\gamma_{1,2}}(x)$:
\begin{equation}
 B^+a^*B^-\H_{n}^{\gamma_{1,2}}(x)=n\sqrt{n}\:B^+\psi_n(x)=n\sqrt{n}\:\H_{n+1}^{\gamma_{1,2}}(x).
\end{equation}
Therefore the creation operator $A^+$ is defined as:
\begin{equation}
 A^+=B^+a^*B^-
\end{equation}
and one also gets $c_1=n\sqrt{n}$. Since $B^+$, $a^*$ and $B^-$ are linear first-order operators
$A^+$ will be a differential operator of third order.

To calculate the annihilation operator $A$, we use the results from (\ref{ec_usar}) to which we apply the annihilation operator $a$:
\begin{equation}
 aB^-\H_{n}^{\gamma_{1,2}}(x)=n\:a\psi_{n-1}(x)=n\sqrt{n-1}\:\psi_{n-2}(x).
\end{equation}
Finally, applying the operator $B^+$, one gets
\begin{equation}
 B^+aB^-\H_{n}^{\gamma_{1,2}}(x)=n\sqrt{n-1}\:B^+\psi_{n-2}(x)=n\sqrt{n-1}\:\H_{n-1}^{\gamma_{1,2}}(x),
\end{equation}
wherefrom we get the form of the $A$ operator as,
\begin{equation}
 A= B^+aB^-.
\end{equation}
Similarly to the operator $A^+$, the operator $A$ is a differential operator of third order and $c_2=n\sqrt{n-1}$.

Similarly to the factorization introduced by Mielnik, the function of subindex zero cannot be obtained using the $A^+$ and $A$ pair only.
In this case, they transform the functions  $\H_{n}^{\gamma_{1,2}}(x)$ into the functions $\H_{n+1}^{\gamma_{1,2}}(x)$
or $\H_{n}^{\gamma_{1,2}}(x)$ into $\H_{n-1}^{\gamma_{1,2}}(x)$ but the constants $c_1$ and $c_2$
are different of those obtained when the operators $a$ and $a^*$ are applied to $\psi_n$.

\section{Can one have other factorizations ?}

One of the obvious questions in the case of the factorization introduced by Mielnik is what happens if instead of requiring
$bb^*=aa^*$ one imposes $b^*b=a^*a$ which looks natural because $aa^*+\dfrac{1}{2}=a^*a-\dfrac{1}{2}=H$.
However, factorizing in this manner leads to $\phi(x)=\left( \gamma-\int_0^xe^{t^2}\:dt
 \right)^{-1}e^{x^2}$, which is a monotonically increasing function which is singular at the particular value of $x$ where the integral equals
$\gamma$. Does the same problem occur for our factorization ?

Suppose now that instead of requiring $B^-B^+=H+\frac{1}{2}$ as defined in (\ref{Bs}) we ask that $B^+B^-=H-\frac{1}{2}$.

We again can get a Riccati equation similar to (\ref{ed0}) but this time for the product $\alpha\beta$. It is then easy to obtain the factorizing coefficients in the following explicit form
\begin{equation}\label{alphabetatwo}
 \alpha_{\kappa_1 \kappa_2}(x) = \pm\sqrt{1+\frac{\kappa_2e^{x^2}}{\left( \kappa_1-\int_0^x e^{t^2}dt \right)^2}}\,\quad
  \beta_{\kappa_1 \kappa_2}(x)=\pm\frac{x\left( \kappa_1-\int_0^x e^{t^2}dt \right)+e^{x^2}}
{\sqrt{\left( \kappa_1-\int_0^x e^{t^2}dt \right)^2+\kappa_2e^{x^2}}}~,
\end{equation}
where $\kappa_1$ and $\kappa_2$ are constants.
However, one can see that there are problems with the continuity of the functions, namely $\kappa_1-\int_0^x e^{t^2}dt$ could be zero for special values of $x$.
For this reason, we present only the particular case
\begin{equation}\label{ab2}
 \alpha(x)\beta(x) = x~.
\end{equation}

Thus,
\begin{equation}
  \alpha_{\gamma_3}(x) = \sqrt{1+\gamma_3e^{x^2}}~,\qquad \beta_{\gamma_3}(x) = \frac{x}{\sqrt{1+\gamma_3e^{x^2}}}.
\end{equation}
To have $B^+$ and $B^-$ well defined along the full $x$ axis we require $\gamma_3\geq 0$.
We notice that in this case there is no particular value for the constant that can lead to the factorization introduced by Mielnik 
or to the $\delta$-parameter non-mutually-adjoint factorization. This happens because we have started with the different condition $B^+B^-=H-\frac{1}{2}$ and we use a particular solution of $\alpha(x)\beta(x)$. However, we can go to the standard factorization by making $\gamma_3=0$.


We now define the operator $\widetilde\L_{\gamma_3}=B^-B^+ -\frac{1}{2}$, or explicitly
\begin{equation}
 \widetilde\L_{\gamma_3} = -\frac{1}{2}\:\frac{d^2}{dx^2}-\frac{\gamma_3xe^{x^2}}{1+\gamma_3e^{x^2}}\:\frac{d}{dx}
+\frac{1 + x^2 + \gamma_3e^{x^2}}{2\left( 1+\gamma_3e^{x^2} \right)^2} - \frac{1}{2} \label{2operador_n}
\end{equation}
and its eigenfunctions $\H_n^{\gamma_3}(x)$ as follows
\begin{equation}\label{ufunction}
\H_n^{\gamma_3}(x)=B^- \psi_{n+1}(x)~.
\end{equation}
Thus
\begin{equation}
\widetilde\L_{\gamma_3}\H_{n}^{\gamma_3}=\left( B^-B^+ -\frac{1}{2} \right)\left( B^- \psi_{n+1} \right)=
B^-\left( B^+B^--\frac{1}{2} \right)\psi_{n+1}=\lambda_n\H_n^{\gamma_3}.
\end{equation}
Differently from the factorization used in the previous section, 
in this case it is not necessary to calculate
$\H_0^{\gamma_3}$ because it is defined in (\ref{ufunction}). Moreover $B^-=\alpha^{-1}_{\gamma_3}(x)
(\: d/dx+x \:)=\alpha^{-1}_{\gamma_3}(x)a$, and therefore the unnormalized eigenfunctions should be
$\H_{n}^{\gamma_3}=\alpha^{-1}_{\gamma_3}(x)\psi_n(x)$. Following the analysis performed in the previous section, 
we will seek the self-adjoint form of the second-order differential operator (\ref{2operador_n}). It is easy to see that we have to multiply by the following factor
\begin{equation}
 -2\:\exp\left( 2\int^x \frac{\gamma_3xe^{x^2}}{1+\gamma_3e^{x^2}}\:dx \right) = -2\alpha^2(x) = -2 \left( 1+\gamma_3e^{x^2} \right).
\end{equation}
Under multiplication, (\ref{2operador_n}) shows that the functions $\H_{n}^{\gamma_3}$ are orthogonal by construction leading to the following eigenvalue problem
\begin{equation}
 \L_{\gamma_3}\H_{n}^{\gamma_3}+\lambda_n\omega_{\gamma_3}(x)\H_{n}^{\gamma_3}=0,
\end{equation}
where
\begin{equation}
 \L_{\gamma_3} = \left( 1+\gamma_3e^{x^2} \right)\frac{d^2}{dx^2} + 2\gamma_3xe^{x^2}\:\frac{d}{dx}
+\frac{\gamma_3e^{x^2}+\gamma_3^2e^{2x^2}-x^2}{ 1+\gamma_3e^{x^2}}
\end{equation}
is the one-parameter self-adjoint operator with the weight function $\omega_{\gamma_3}(x) = 2 \left( 1+\gamma_3e^{x^2} \right)$ which is isospectral to the quantum harmonic oscillator operator, obtained in the limit $\gamma_3\rightarrow 0$. Interestingly, in the large limit $\gamma_3>1$,
one can obtain the following differential equation
\begin{equation}
\left[ \frac{d^2~}{dx^2} + 2x \frac{d~}{dx} +2(n+1) \right] G_n(x)=0~,
\end{equation}
which differs from the Hermite equation only in the sign in front of the first derivative. The corresponding eigenfunctions are of the quantum oscillator type,
but with the Gaussian factor of double width, i.e., $G_n(x) = c_n e^{-x^2} H_n(x)$.

\section{Conclusion}

This work contains the generalization of the factorization procedure introduced by Mielnik for the quantum harmonic oscillator performed with
a pair of non-mutually adjoint factorization operators. It is also an extension of the previous paper \cite{1} and leads to a two-parameter self-adjoint second-order operator which contains the standard harmonic oscillators, the one-parameter harmonic oscillators introduced by Mielnik, and even the Hermite operator for special values of its parameters. Both parameters are essentially integration constants, one coming from the Riccati equation and the other from the Bernoulli equation which occur in the factorization procedure. The Riccati parameter is a shift parameter controlling the position of the maxima of the eigenfunctions along the $x$-axis, while the Bernoulli parameter is related to the shape of the eigenfunctions. In the final part, we have also introduced a Bernoulli parameter case which is different of the one discussed in \cite{1}. Moreover, these self-adjoint operators are of the effective mass Schr\"odinger type with position-dependent mass which are known to have important applications \cite{GN}.

\bigskip

{\small

{\bf Acknowledgement}:
The first author wishes to thank CONACyT for the master program fellowship. The third author thanks CONACyT for a sabbatical fellowship.


}

\end{document}